\documentclass[10pt,preprint2]{aastex} 		

\def\mpchi{\,h^{-1}{\rm {Mpc}}}
\def\kpchi{\,h^{-1}{\rm {kpc}}}
\def\kms{\,{\rm {km\, s^{-1}}}}

\def\msunhi{\,h^{-1}{\rm M_\odot}}

\begin{document}

\title{A fitting formula for the merger timescale of galaxies in
hierarchical clustering }

\author{C. Y. Jiang$^{1,2}$, Y. P. Jing$^{1}$, A. Faltenbacher$^{1}$, W. P. Lin$^{1}$, Cheng Li$^{1}$}
\affil{$^1$Shanghai Astronomical Observatory, Nandan Road 80, Shanghai, China}
\affil{$^2$
Graduate School of the Chinese Academy of Sciences, 19A, Yuquan Road, Beijing, China}

\begin{abstract}
We study galaxy mergers using a high-resolution cosmological
hydro/N-body simulation with star formation, and compare the measured
merger timescales with theoretical predictions based on the
Chandrasekhar formula. In contrast to Navarro et al., our numerical
results indicate, that the commonly used equation for the merger
timescale given by Lacey and Cole, systematically underestimates the
merger timescales for minor mergers and overestimates those for major
mergers. This behavior is partly explained by the poor performance of
their expression for the Coulomb logarithm, $\ln (m_{\rm pri}/m_{\rm sat})$.
The two alternative forms $\ln (1+m_{\rm pri}/m_{\rm sat})$ and
$1/2\ln [1+(m_{\rm pri}/m_{\rm sat})^2]$ for the Coulomb logarithm 
can account for the mass dependence of merger timescale successfully, but
both of them underestimate the merger time scale by a factor 2. 
Since $\ln (1+m_{\rm pri}/m_{\rm sat})$ represents the mass dependence
slightly better we adopt this expression for the Coulomb logarithm.
Furthermore, we find that the dependence of the merger timescale on
the circularity parameter $\epsilon$ is much weaker than the widely
adopted power-law $\epsilon^{0.78}$, whereas
$0.94{\epsilon}^{0.60}+0.60$ provides a good match to the data. Based
on these findings, we present an accurate and convenient fitting
formula for the merger timescale of galaxies in cold dark matter models.
\end{abstract}


\section{Introduction}

Dynamical friction plays a crucial role in the formation and evolution
of galaxies. During the merger of two dark matter halos, galaxies in a
less massive halo will become the satellite galaxies of the more
massive one. These satellite galaxies gradually lose their energy and
angular momentum under the action of dynamical friction and are
predestined to sink to the center of the massive dark matter halo, if
they are not disrupted by the tidal force. 

Dynamical friction takes effect through interaction of galaxies with
background dark matter particles. \cite{chandrasekhar43} gave a
description for this phenomenon for an idealized case where a rigid
object moves through a uniform sea of collisionless matter particles. This
description can be applied to the case of a satellite galaxy moving in
a dark matter halo. The orbits of dark matter are deflected by the
galaxy, which produces an enhancement of dark matter density behind
the galaxy.  Consequently, the galaxy suffers a steady deceleration by the
drag of the wake, and will eventually merge to the central galaxy of the dark
matter halo. The merger timescale, i.e. the time elapsing between
entering the virial radius of the dark matter halo and final
coalescence of satellite and central galaxy, can be derived using 
Chandrasekhar's formula \citep[see, e.g.,][]{binney87}. Additionally,
taking into account the dependence on the orbital circularity
\cite{lacey93} derived the following expression for the merger
timescale of a satellite galaxy orbiting around a massive halo with
circular velocity $V_{\rm c}$
\begin{equation}
T_{\rm Chandra}=\frac {1} {2} \frac {f(\epsilon)V_{\rm c}r_{\rm c}^2}
{C G m_{\rm sat}\ln\Lambda}\ ,
\label{eq:dynf}
\end{equation}
where $\epsilon$ is the circularity parameter of the satellite's orbit
and $r_{\rm c}$ is the radius of a circular orbit with the same energy
as the satellite's orbit. $f(\epsilon)$ describes the dependence of
$T_{\rm Chandra}$ on the orbital circularity, and is approximated by
$f(\epsilon) \sim \epsilon^{0.78}$ for $\epsilon> 0.02$
\citep{lacey93}.  $C$ is a constant, approximately equal to 0.43, and
$m_{\rm sat}$ is the satellite mass. $\ln \Lambda$ is the Coulomb
logarithm, which is given $\ln(d_{\rm max}/d_{\rm min})$,
where $d_{\rm max}$ is the maximum relevant impact parameter at which
background particles are scattered into the wake and $d_{\rm min}$ is
the minimum impact parameter \citep{chandrasekhar43,white76}. It is
expected to be applicable for cases where the satellite mass is much
smaller than that of the primary halo. 

There have been many works which used N-body simulations to check the
validity of Chandrasekhar's formula and its application to the merging
of satellite and central galaxies, but no consensus has been 
reached on the accuracy of such applications. This is because a galaxy
merger is a more complicated process than a pure motion of a rigid
body through an uniform collisionless matter distribution  as
considered by Chandrasekhar. The primary halo has a density increasing
inward to the halo center, which makes it nontrivial to choose the
maximum impact parameter for the Coulomb logarithm \citep{hashimoto,jb00,
vdbosch99}. Because the satellites lose their mass due to the 
tidal interaction by the primary halo, one has to follow both the trajectory
and the mass evolution of the satellites to derive their merger
timescale. Unfortunately, there is still a considerable amount of
uncertainties in modeling these processes
\citep{tormen98,gao04,zentner05}. A further complication is that due to 
the similar orbits of the tidally stripped mass and the satellite
itself the tidal debris will trail the satellite for a significant
amount of time which in turn will exert a drag force on the 
satellite \citep{fujii,fellhauer07}. Besides, the merger can 
alter the structure of the primary halo which is another complication for
accurately computing the merger timescale \citep{zaritsky88,cora97}.

It is however very useful to give a simple prescription for the merger
timescale of the satellites. \cite{navarro95} used an
N-body/hydrodynamics simulation with gas cooling to determine the 
merger time scales. Their simulation didn't include a recipe for
star formation, thus they used the cold gas at the cores of dark
matter halos as a proxy for galaxies. They found a good agreement with
the prediction of equation (\ref{eq:dynf}) if the satellite mass
$m_{\rm sat}$ is taken to be the sum of the cold gas core and the
associated dark matter halo at the moment when it crosses the virial
radius of the primary halo for the first time. They further pointed
out that the predicted merger timescale is too long if only the cold
gas is taken for the satellite mass. 

The N-body study of Navarro et al. provides a strong support for using
equation (1) to determine the merger timescale in both theoretical and
observational studies, if $m_{\rm sat}$ is taken to be the total mass
of the satellite at the virial radius of the primary halo. For
example, this equation is an important ingredient in modeling mergers
of galaxies in analytical studies of galaxy formation
\citep[e.g.,][for an excellent review] {kauffmann99, cole00,monaco00,
somerville99, menci02, nagashima2002, hatton03,K007, baugh06} and in
understanding the merger rates of galaxies in the cosmological context
\citep[e.g.][]{ot79,lin04,gill05,maller06,chw07,zheng071,zheng072}.
However, there are indications that the Navarro et al. prescription
underestimates the merger time or overestimates the merger
rate. \cite{springel01} and \cite{kang05} found that the luminosity of
central galaxies in rich clusters is reduced if the orbital evolution
of satellites is determined by high-resolution N-body simulations
compared to the luminosities based on the Navarro et al. merger
rates. We also note that the N-body experiment by \cite{colpi99} gave
a merger timescale which is longer than what Navarro et
al. suggests. They even found an much weaker dependence on the
circularity with the exponent only about 0.4 (instead of
0.78). Therefore it is not yet clear what causes these discrepancies,
especially the one between Colpi et al. and Navarro et al.. It would
be helpful to point out that Navarro et al. used a cosmological
hydro/N-body simulation with gas cooling and included both major and
minor mergers in their study, while Colpi et al. used N-body
simulations of galaxy mergers and considered minor mergers only.

In this paper, we will use a high-resolution hydro/N-body cosmological
simulation to clarify this situation. In the simulation, gas cooling
and star formation are included so that the galaxy mergers can be
identified unambiguously and the merger timescale can be well
measured. Our results can be directly compared with Navarro et al.,
therefore, they will be used to study the origin of the
discrepancies mentioned above. We will show that the Navarro et
al. prescription actually underestimates the merger time for minor
merges, qualitatively in good agreement with Colpi et al., but
overestimates it for major mergers. In light of our 
simulation results, we will propose an accurate fitting formula for
the merger timescale that accounts well for the dependences on
mass and circularity of the individual satellites, and can therefore
accommodate both, minor and major merger events.

The paper is organized as follows. In Section 2, we describe our
simulation and our method for calculating the merger timescale in the
simulation. Section 3 gives a comparison between our simulation result
and the theoretical prediction. A new fitting formula for the merger
timescale is derived in Section 4. Finally, we summarize our results
in Section 5.

\section{Merger timescales in simulation}

\subsection{The simulation}

A parallel version of the SPH code GADGET2
\citep{springel01,springel05} is used to simulate the structure formation
and evolution in the Universe.  The cosmological parameters we use are
$\Omega _{\Lambda}=0.732$, $\Omega _{\rm m}=0.268$, $\Omega _{\rm
b}$=0.044, $\sigma_{\rm 8}$=0.85, and a Hubble constant $H_{\rm
0}=100h \kms{\rm Mpc}^{-1}$ with $h=0.71$.  The box is $100\mpchi$ on
a side, with $512^{3}$ dark matter particles and $512^{3}$ gas
particles. The resulting mass resolution for dark matter and gas
particles is $4.6 \times 10^8 \msunhi$ and $9.2 \times 10^7 \msunhi$,
respectively. The simulation includes the physical processes of
radiative cooling and star formation.  It also includes supernova
feedback, outflows by galactic winds, and a sub-resolution multiphase
model for the interstellar medium as detailed in \citet{sh03}. The
simulation has the same mass resolution and model parameters as the
star formation run of \citet{jing06}, except that the softening length
of the gravitational force is greatly reduced in the current
simulation, where we use a spline kernel \citep{springel05}, roughly
equivalent to a Plummer force softening of $4.5\kpchi$ (comoving). 
There are a total of 88 snapshot outputs from $z = 2.0$ to the
present time, $z=0$, with an equal logarithmic scale factor interval 
of $\Delta \ln a=0.01$ between two consecutive outputs. The large 
number of outputs enables us to accurately sample orbits of
satellites within massive halos. Both, the good force resolution and 
the dense sampling of snapshots, are crucial for the current study.

\subsection{Construction of halo merger trees}

Dark matter halos are identified using the friends-of-friends (FOF)
method, with a linking length of 0.2 times the mean inter-particle
separation. To obtain a sufficient number of halos with reasonable mass
resolution, we only focus on the halos with masses $m_{\rm vir}> 5 \times
10^{12}\msunhi $ at the present epoch. The virial mass of a halo
$m_{\rm vir}$ is defined as the mass enclosed within the virial radius
$r_{\rm vir}$ within which the mean mass density is $\Delta(z)$ times
the critical density of the universe at redshift z. For $\Delta(z)$ we
adopt the fitting formula for flat universes provided by \cite{bryan98},
\begin{equation}
\Delta(z)=18\pi^{2}-82x-39x^{2}\,,
\end{equation}
where $x$ is the density parameter for the vacuum density (the
cosmological constant) at redshift $z$. 

Then we trace these halos back to $z=2.0$ to construct the main
branch of the merger tree for each halo. For halo {\it A} at some
snapshot, halo {\it B} at an earlier snapshot which, among all
its progenitors, contributes the largest number of particles to {\it
A} is defined as the main progenitor of {\it A}. All the other
progenitors of halo {\it A}, each of which is required to have more than
half its particles merging with {\it A}, are taken as satellite halos
of halo {\it B}, while {\it B} is called the primary halo. Note, we
use 'satellite' to represent the whole halo, including both, dark
and stellar matter.  

We do not use the orbital energy as the criterion to identify a
satellite as being bound or unbound, since an orbit that starts out
unbound will not necessarily remain unbound, because dynamical
friction may sufficiently reduce its energy, see e.g.,
\cite{benson05}. They find that only about 2\% of all initially
unbound orbits fail to become bound and so escape from their primary
halo. Furthermore, to reduce artificial effects caused by the finite
numerical resolution we keep only those satellites that have central
galaxies more massive than $ 2.0\times 10^{10}\msunhi$. Typically,
these  satellite galaxies are surrounded by a dark matter halo
comprising  more than 1000 particles before entering the primary halo.

\subsection{Merging timescale of galaxies}

The galaxies are also identified with the friends-of-friends method
applied to the star particles but with a small linking length of
$4.88\kpchi$. Besides the central galaxies of the primary halos we
only focus on those galaxies which have been the central galaxies of 
the approaching satellite halos. Throughout, the former will be
referred to as {\it central} and the latter as {\it satellite}
galaxies. The descendant of a galaxy, {\it C}, is the galaxy in a
subsequent snapshot which shares the most star particles with {\it
C}. A galaxy merger is identified if the satellite galaxy and the
central galaxy begin to have the same descendant at one snapshot, and
continue to have the same descendant for the following four snapshots
($\geq$ half of the dynamical time of a halo). We use this criterion to
ensure that the merger is a real merger not just a close flyby.

The merger timescale is defined as the time elapsed between the moment
when the satellite galaxy first crosses the virial radius of the
primary halo and the final coalescence of satellite and central galaxy.
The computation of the merger timescale from the simulation involves
four distinct snapshots: snapshot {\it a}, the last snapshot  
for which the satellite halo is identified as a single halo; 
snapshots {\it b} and {\it b+1},  between snapshot {\it b} and
snapshot {\it b+1} the satellite galaxy crosses the virial radius of
the primary halo for the first time; and finally, snapshot {\it c}, 
beginning of the coalescence of satellite and central galaxy.
To accurately determine the point in time when the satellite galaxy
enters the primary halo, we assume that the satellite galaxy moves
with constant velocity from its actual location, both measured at
snapshot {\it b}, until it hits the virial radius, which has been
fixed at snapshot {\it a}. However, a substantial fraction
($\sim14\%$) of the satellite galaxies do not reach the virial radius
within the time interval between snapshot {\it b} and {\it b+1}.
This happens because, in general, satellites are in accelerated
motion. In such cases, we choose snapshot {\it b+1} as the time at
which the satellite reaches the virial radius. Due to the dense time
sampling by the large number of snapshots this uncertainty constitutes
only a marginal source of error. 

Finally, the merger timescale for each completed merger event is
defined to be the interval between the time when the satellite first
enters the virial radius and the middle point between snapshots {\it
c} and {\it c-1}. 

\begin{figure}
\begin{center}
\plotone{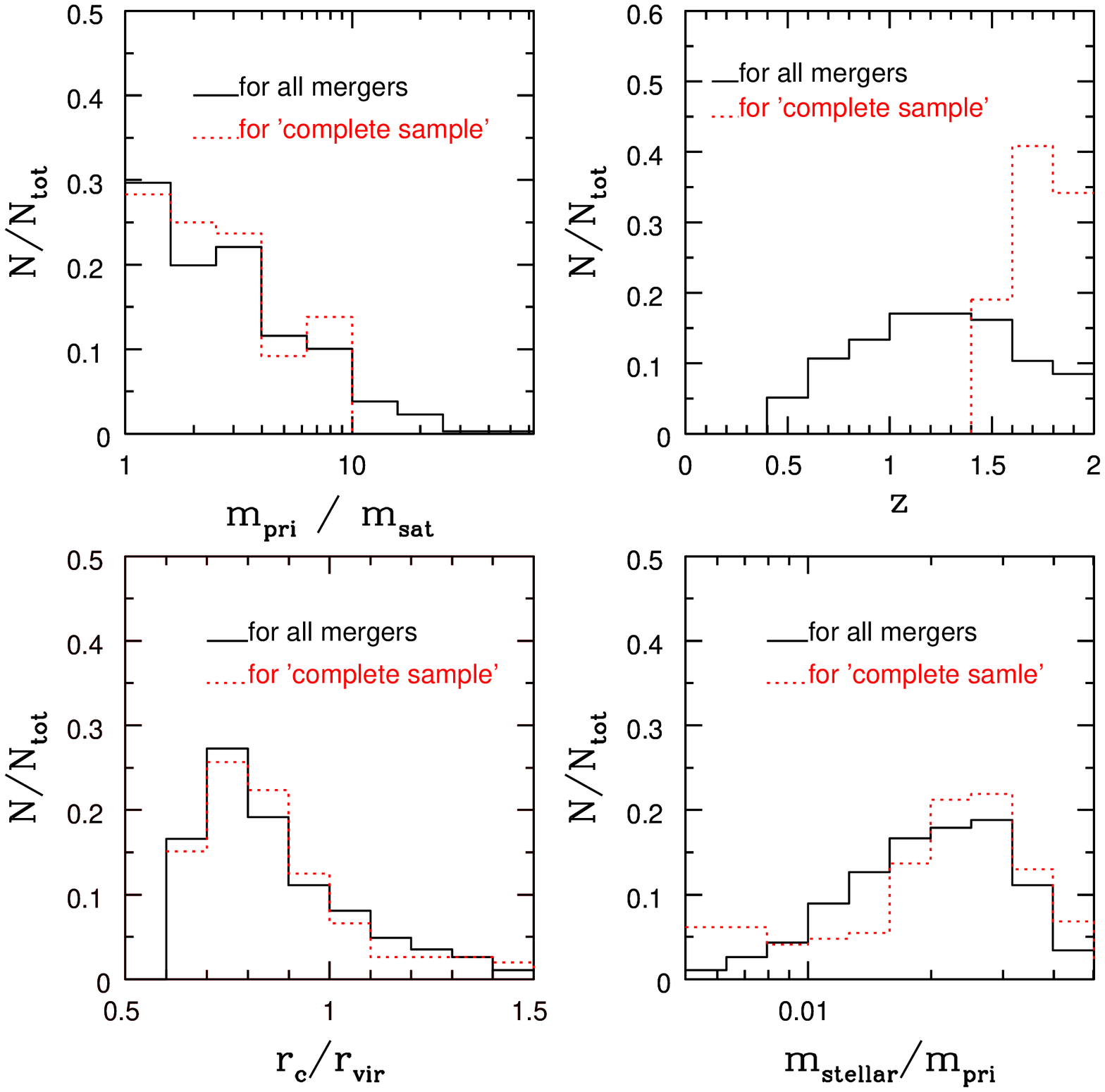}
\caption{\label{fig:dis} Basic characteristics of all mergers (solid
histograms) and the mergers in the complete sample (dashed
histograms). The panels from the top left to the bottom left clockwise
show the distributions of the mass ratio of the primary halo to the
satellite, the first crossing redshift, the mass ratio of the
stellar mass of the central galaxy to the primary, and the ratio
of $r_c$ to the virial radius $r_{\rm vir}$ respectively.}   
\end{center}
\end{figure}

Some basic statistical properties of the mergers are presented in
Figure \ref{fig:dis}. There is almost an equal amount of major mergers
and minor mergers, if we use the mass ratio $m_{\rm pri}/m_{\rm
sat}=3$ as the dividing line. Since we examine only snapshots starting
from redshift $z=2$, the redshifts, at which the eventually merging
satellites first cross the virial radius of the primary halo, span the
range between $z=0.4$ and $z=2$. (Satellites which approach more
recently than $z=0.4$ do not have sufficient time to merge with the
central galaxy.)  The ratio of the stellar mass of a central galaxy to
the dark matter mass of the primary halo varies from 0.5\% to 5\% with
an average of 2\%. This ratio is in reasonable agreement with the
observed values of galaxy groups\citep{g071,lin03}. The satellite
sample has a wide spectrum of orbital
energies, as displayed by the distribution of $r_c/r_{\rm vir}$, which
ranges from $0.6$ to $1.5$ with an average 0.8. Thus, we believe that
our sample represents a typical sample of galaxy mergers.

\section{Comparison with theory}

Equation (\ref{eq:dynf}) is only applicable for mergers with
mass ratios $m_{\rm pri}/m_{\rm sat} \gg 1$, where $m_{\rm pri}$ and
$m_{\rm sat}$ stand for the mass of the primary and the satellite
halo, respectively \citep{binney87}. As mentioned above $\Lambda$ in
the Coulomb logarithm $\ln\Lambda$ is defined as the ratio between
maximal and the minimal impact parameters ($d_{\rm max}/d_{\rm min}$)
for which encounters between the satellite and the dark matter
particles can be considered effective . An equivalent expression for 
$\Lambda$ is given by 
\begin{equation}
\label{eq:coulomb}
\Lambda \equiv \frac {d_{\rm max} {V_{\rm typ}}^{2}}{G(m_{\rm sat}+m_{\rm dm})}
= \frac {m_{\rm pri}}{m_{\rm sat}}\ ,
\end{equation}
where $V_{\rm typ}$ and $m_{\rm dm}$ are the typical velocity and mass
of background dark matter particles. The transition from the middle to
the expression on the right hand side is obtained by setting $d_{\rm
max}=r_{\rm pri}$ (the radius of the primary halo), $V_{\rm typ}
\approx V_{\rm pri}$ (the circular velocity of the primary halo), and
assuming $m_{\rm dm}\ll m_{\rm sat}$. 

Therefore,  according to equation (\ref{eq:dynf}) a correct estimate of
the satellite mass is pivotal for the determination of the dynamical
friction timescale. A satellite orbiting in the potential well of the
primary halo loses a large fraction of its initial mass due to the
exposure to the global tidal field \citep[e.g.,][]{tormen98,gao04,shaw07}
and due to high-speed encounters with other satellites
\citep[e.g.,][]{moore96, gnedin03}. Based on a hydro/N-body simulation
\cite{navarro95} investigated the dependence of the dynamical friction
time scale on the the Coulomb logarithm $\ln \Lambda=\ln (m_{\rm
pri}/m_{\rm sat})$ by considering two extreme choices for $m_{\rm
sat}$: (1) $m_{\rm sat}$ was considered to be the total virial mass of
the satellite before entering the primary halo, i.e. the sum of the
gas (representative for the stellar component in their simulation) and
the cold dark matter within the satellite's virial radius,  
(2) $m_{\rm sat}$ only accounted for the cold gas associated with the
satellite galaxy at the center of the approaching dark matter halo.
They found when the total virial mass is chosen for $m_{\rm sat}$
equation (\ref{eq:dynf}) gives a good prediction for the merger time
scale although the scatter is very large. If only the cold gas is
adopted for $m_{\rm sat}$ equation (\ref{eq:dynf}) significantly
overestimates the merger timescale because the cold gas mass is always
much smaller than the virial mass. Based on this numerical
investigation, equation (\ref{eq:dynf}) with the initial satellite
virial mass for $m_{\rm sat}$ is widely used in galaxy formation
studies \citep[e.g.,][]{cole00,kauffmann99,kang05}. Here, we also
follow this convention for $m_{\rm sat}$. 

\begin{figure}
\begin{center}
\plotone{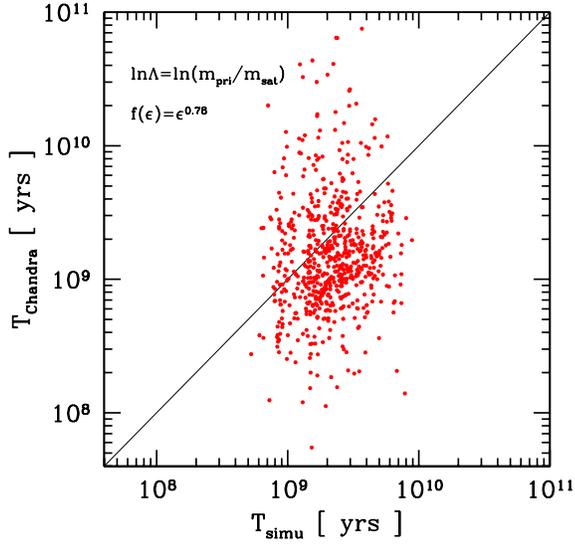}
\caption{\label{fig:merg_all_navarro_rc_1} Comparison of the merger
timescale $T_{\rm simu}$ in the simulation with theoretical
dynamical friction timescale $T_{\rm Chandra}$ from equation
$(1)$. The solid line is $T_{\rm Chandra}=T_{\rm simu}$.  The
Coulomb logarithm is in the form $\ln\Lambda=\ln(m_{\rm vir}/m_{\rm
sat})$, and $f(\epsilon)={\epsilon}^{0.78}$.}
\end{center}
\end{figure}

With the present analysis we aim to examine the validity of equation
(\ref{eq:dynf}) by means of a cosmological high resolution N-body/hydro
simulation. Figure ~\ref{fig:merg_all_navarro_rc_1} compares the
merger timescale $T_{\rm Chandra}$ computed according to equation  
(\ref{eq:dynf}) with the merging time $T_{\rm simu}$ measured in the
simulation. The solid diagonal displays $T_{\rm Chandra}=T_{\rm
simu}$.  The results indicate a qualitative agreement between the
prediction of equation (\ref{eq:dynf}) and the time scales measured
from the simulation. However, scatter between $T_{\rm Chandra}$ and
$T_{\rm simu}$ is extremely large.  To see whether the large scatter
is caused by the failure of equation (\ref{eq:dynf}) for mass ratios
$m_{\rm sat}/m_{\rm pri}\approx 1$ we plot the median value of $T_{\rm
simu}/T_{\rm Chandra}$ as a function of $m_{\rm sat}/m_{\rm pri}$ in
Figure~\ref{fig:mass_dpd_all_2} (the solid line).  The figure clearly
shows that the time ratio increases monotonically with decreasing mass
ratios. That is, the time ratio is significantly smaller than 1 (0.55
for the mass ratio larger than 0.65) for the major mergers and
approaches 4 for minor mergers (for the mass ratio smaller than
0.065). This implies that equation (\ref{eq:dynf}), which is expected 
to be valid for minor mergers, actually underestimates the merger time
scale for them. This result is in approximate agreement with
\cite{colpi99} who found that the friction timescale for $m_{\rm
sat}/m_{\rm pri}\approx 0.02$ is underestimated by a factor of $2$ if 
equation (\ref{eq:dynf}) is used.  On the other side, our result
points out that equation (1) significantly overestimates the dynamical
friction time scales for major mergers. Despite the fact that the
formula is not expected to be applicable to major mergers, it is still
widely used for major mergers in the literature.
Our findings do not agree with \cite{navarro95} who advocate a good
agreement between their simulation result and equation (1) for 
minor mergers with mass ratios less than $0.5$. In the next
section, we will use our simulation data to improve the description
for the merger timescale in the hierarchical clustering scenario. 

\section{Fitting formula for the merger timescale in cosmological
context}
\label{sec:fitting}

First, because $r_{\rm c}\approx r_{\rm vir}$ we rewrite the formula
of $T_{\rm Chandra}$ as
\begin{equation}
\label{eq:dfrc}
T_{\rm Chandra}=\frac {1}{2} \frac{f(\epsilon)}{C}\frac {m_{\rm
  pri}}{m_{\rm sat}} \frac {1}{\ln\Lambda}\frac{r_{\rm c}}{V_{\rm c}}
  \end{equation}
where ${r_{\rm c}}/{V_{\rm c}} \propto 1/\sqrt{G\rho}$ and $\rho$ is
the mean mass density of the halo at that redshift. Thus ${r_{\rm
c}}/{V_{\rm c}}$ is proportional to the age of the Universe at the
epoch being considered\footnote{$r_c$ is about $r_{\rm vir}$ but
there is scatter, so the statement is valid approximately},
independent of primary and/or satellite halo masses. Consequently, the
mass dependence of $T_{\rm Chandra}$ is solely accounted for by the
mass ratio between satellite and primary halo, and its circularity
dependence is included by the function $f(\epsilon)$. It is suggesting
to isolate those two dependencies to find the cause of the
discrepancies between the merger time scales derived from 
equation (\ref{eq:dynf}) and the simulation. Therefore, in the
following section we will focus on the dependence of the merging time
scales on the mass ratios. Subsequently, we will examine the
circularity dependence in detail. Finally, these investigations will
lead us to a new description of merger time scales in the cosmological
context. 

\subsection{Dependence on the mass ratio and Coulomb logarithm}

\begin{figure}
\begin{center}
\plotone{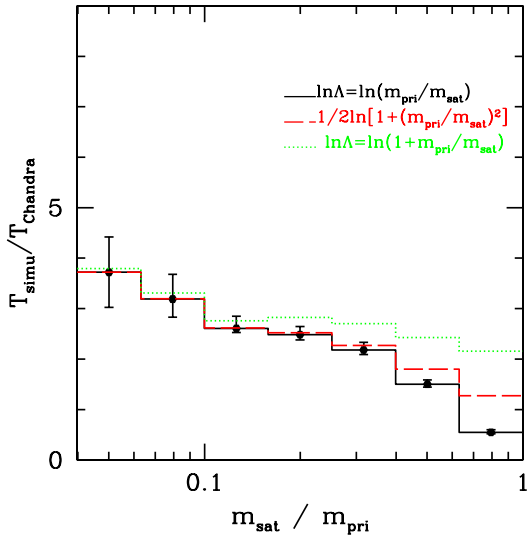}
\caption{\label{fig:mass_dpd_all_2} Mass dependence of the median value of 
$T_{\rm simu}/T_{\rm Chandra}$, when different forms are used for the Coulomb logarithm.}
\end{center}
\end{figure}

The strong dependence of $T_{\rm simu}/T_{\rm Chandra}$ on the mass ratio
$m_{\rm sat}/m_{\rm pri}$ shown in Figure~\ref{fig:mass_dpd_all_2} (the
black solid line) indicates that the mass dependence of $T_{\rm
Chandra}$ as described by equation $(1)$ is incorrect. Here, we first
consider to revise the Coulomb logarithm. In fact, in the original
derivation of the formula \citep[see,][]{binney87}, the Coulomb
logarithm should read as $\frac{1}{2}\ln (1+\Lambda^2)$. Only if the
satellite mass is much smaller than the primary mass this expression
can be written as $\ln \Lambda$. In the literature $\frac{1}{2}\ln
(1+\Lambda^2)$ is simply used to include mergers that do not satisfy
the condition $m_{\rm sat}\ll m_{\rm pri}$
\citep[e.g.][]{somerville99}.  But another version, namely
$\ln(1+m_{\rm pri}/m_{\rm sat})$, is even more widely used for the
same purpose \citep[e.g.][]{springel01b,volonteri03,kang05} despite
the fact that there is no clear physical motivation for adopting
it. Here we examine the mass dependence using these two alternative
forms for the Coulomb logarithm. 

The red dashed line and the green dotted line in
Figure~\ref{fig:mass_dpd_all_2} show the mass dependence of $T_{\rm 
simu}/T_{\rm Chandra}$ for these two alternative forms of the Coulomb
logarithm. For mass ratios less than 0.1 the two curves are quite
similar to that of $\ln \Lambda$ (solid line). But, for  mass
ratios $\sim 1$ they display substantial differences.
The mass dependence becomes significantly smaller for these two forms,
especially for the form $\ln(1+m_{\rm pri}/m_{\rm sat})$, however, it
does not disappear completely.
 
\begin{figure}
\begin{center}
\plotone{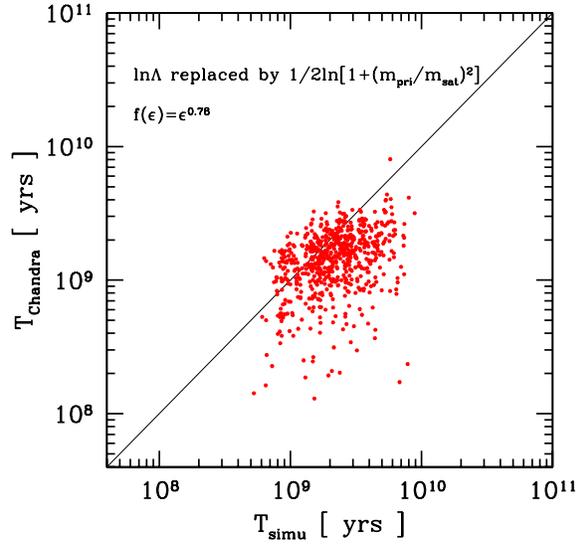}
\caption{\label{fig:merg_all_ori_4} The same as Figure 1, but we replace
$\ln\Lambda$ by $\frac{1}{2}\ln(1+{\Lambda}^{2})$, while $\Lambda$ is
unchanged, and replace $r_{\rm c}$ by $r_{\rm vir}$.}
\end{center}
\end{figure}

\begin{figure}
\begin{center}
\plotone{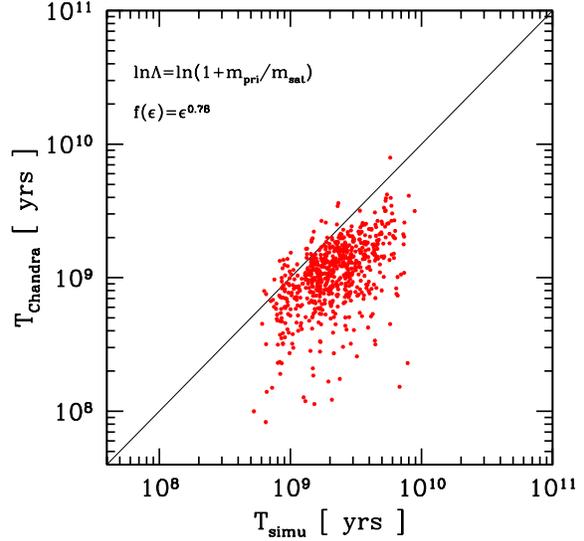}
\caption{\label{fig:merg_all_kang_5} The same as Figure 1, but we replace
$\ln\Lambda$ by $\ln(1+{\Lambda})$, and replace $r_{\rm c}$ by $r_{\rm
vir}$.}
\end{center}
\end{figure}

\begin{figure}
\begin{center}
\plotone{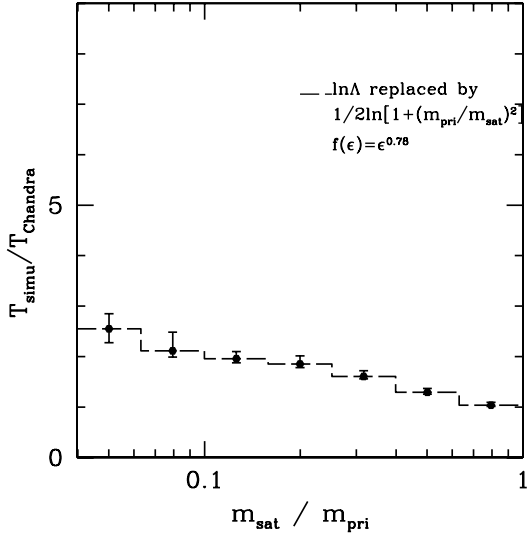}
\caption{\label{fig:mass_dpd_ori_6} The ratio of $T_{\rm simu}$ to $T_{\rm
Chandra}$ as a function of mass ratio for the merger points in Figure 4.}
\end{center}
\end{figure}

\begin{figure}
\begin{center}
\plotone{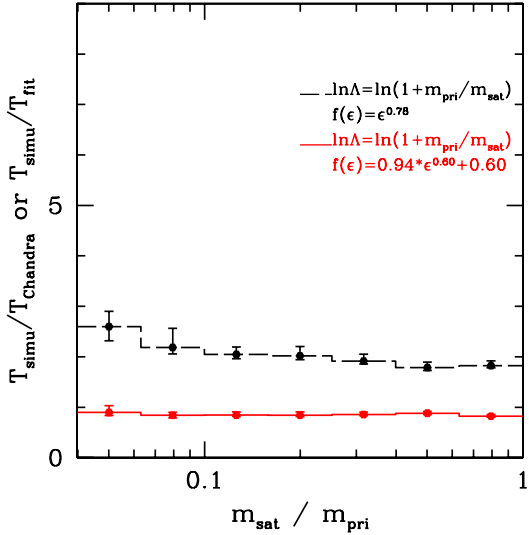}
\caption{\label{fig:mass_dpd_kang_c1_7} The ratio of $T_{\rm simu}$ to $T_{\rm
Chandra}$ as a function of mass ratio for merger points in Figure 5 (the upper 
dashed line) and for those after applying equation (\ref{eq:new}) (the lower solid line).}
\end{center}
\end{figure}

As a trial to improve the description for the mass dependence, we
replace $r_{\rm c}$ in equation (\ref{eq:dynf}) by $r_{\rm vir}$ for
the two forms of the Coulomb logarithm mentioned above.
Figures~\ref{fig:merg_all_ori_4} and~\ref{fig:merg_all_kang_5} show
$T_{\rm simu}/T_{\rm Chandra}$ for $\frac{1}{2}\ln (1+\Lambda^2)$ and 
 $\ln(1+m_{\rm pri}/m_{\rm sat})$, respectively. The plot based on
either of the two forms does not differ much. Here, we want to emphasize
two points.  First, the scatter in the plots is much smaller than in
Figure~\ref{fig:merg_all_navarro_rc_1}. Second, the value of $T_{\rm
Chandra}$ is systematically smaller than that of $T_{\rm simu}$.   
Although the scatter is smaller, it nevertheless will provide
some deeper insight to examine if the scatter depends on the mass
ratio. In analogy to Figure~\ref{fig:mass_dpd_all_2}, we plot in the 
Figures~\ref{fig:mass_dpd_ori_6} and ~\ref{fig:mass_dpd_kang_c1_7}
the median value of $T_{\rm simu}/T_{\rm Chandra}$ as a function
of the mass ratio, for $\frac{1}{2}\ln (1+\Lambda^2)$ and 
$\ln(1+m_{\rm pri}/m_{\rm sat})$, respectively. While there is a
moderate dependence on the mass ratio when the form $1/2\ln[1+(m_{\rm
pri}/m_{\rm sat})^2]$ is used, it is very interesting to recognize
that the dependence of $T_{\rm simu}/T_{\rm Chandra}$ on the mass
ratio for $\ln(1+m_{\rm pri}/m_{\rm sat})$ is strongly reduced.
This implies that the mass dependence of the merger time scale can be
well represented by the form $\ln(1+m_{\rm pri}/m_{\rm sat})$, though
many previous works using this form actually systematically
underestimate the merger time scale or overestimate the merger rate by
a factor 2. 

In the following discussion, we will always use the form $\ln(1+m_{\rm
pri}/m_{\rm sat})$ for the Coulomb logarithm. We prefer to use this
form with $r_{\rm c}$  replaced by $r_{\rm vir}$ as this gives a much
tighter correlation between $T_{\rm Chandra}$ and $T_{\rm simu}$ and
can effectively 
absorb the dependence on the mass ratio. Moreover, in many practical
applications, it is usually easier to use $r_{\rm vir}$ than to use
$r_{\rm c}$.

\subsection{Dependence on circularity and the revised form of $f(\epsilon)$}
\label{sec:revised_fe}

\begin{figure}
\begin{center}
\plotone{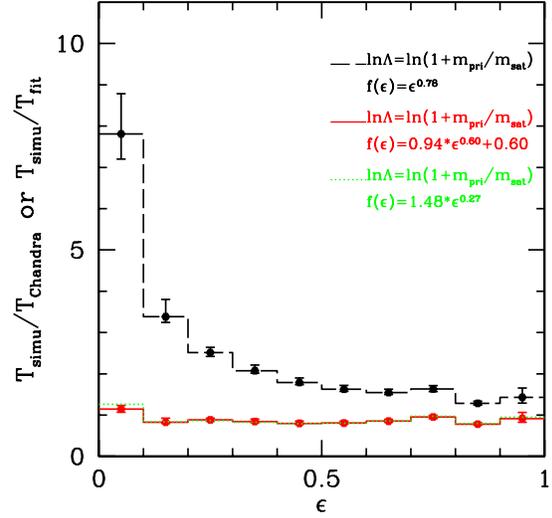}
\caption{\label{fig:circu_dpd_kang_c1_8} The ratio of $T_{\rm simu}$
to $T_{\rm Chandra}$ as a function of $\epsilon$ for merger points in
Figure 5 (the upper dashed line) and for those after applying equation
(\ref{eq:new}) (the lower solid line).  The lower dotted line is for
$f(\epsilon)=1.48*{\epsilon}^{0.27}$ with equation (\ref{eq:new}).}
\end{center}
\end{figure}

Now we check the dependence of the merger time on the initial
circularity parameter $\epsilon$. This parameter is determined from
the velocity and position of a satellite when it first crosses the
virial radius of the primary halo. As in the literature, we assume
that the halo is an isothermal sphere when determining the
circularity. In Figure~\ref{fig:circu_dpd_kang_c1_8} (upper line), we
show the median value of $T_{\rm simu}/T_{\rm Chandra}$ as a function of
circularity, where we have used $\ln(1+m_{\rm pri}/ m_{\rm sat})$ for
the Coulomb logarithm and $f(\epsilon)=\epsilon^{0.78}$ when we
calculate $T_{\rm Chandra}$. The figure shows that the satellites on
very eccentric orbits tend to merge in a much longer timescale
compared to the theoretical prediction. If we still use an exponential
form to represent $f(\epsilon)= \epsilon^\alpha$, the exponent
$\alpha$ should be smaller than the widely used value $0.78$ advocated
by \cite{lacey93}.

Here we explore the form of $f(\epsilon)$ as a function of initial
circularity $\epsilon$. Substituting the merging time in equation
(\ref{eq:dynf}) with what we measure in the simulation, $r_{\rm c}$
with $r_{\rm vir}$, and the Coulomb logarithm with $\ln(1+m_{\rm
pri}/m_{\rm sat})$, we obtain the values of $f(\epsilon)$ for each
merged satellite. Subsequently, we pick the median value of
$f(\epsilon)$ in each circularity bin in our merged satellite
sample. Computing the median value, however, demands some caution.
Because there is considerable scatter in $T_{\rm simu}$ even for the
same circularity and the same mass ratio (which owes to fact that
internal structure and merger history of the primary halo may
introduce some scatter into the merger time scale), there may exist a
selection (or incompleteness) bias against those satellites of long
$T_{\rm merger}$. Those mergers would happen after our fifth last
snapshot and thus be missed in our study. This effect gets more severe
at larger $\epsilon$, because the merger times become systematically
longer on more circular orbits. As a result, only those mergers with
smaller $T_{\rm merger}$ (for the same $\epsilon$) are selected into
the merger sample, which will artificially lower the estimate of
$f(\epsilon)$ for large $\epsilon$. In order to avoid such selection
bias for the determination of $f(\epsilon)$, we construct a complete
merger sample of primary halos and satellites at the first 14
snapshots (redshift $1.55 \sim 2.0$) with mass ratio greater than 0.1
(152 pairs). In this sample, all these satellites but 2 are found to
have merged with the central galaxies of the primary halos before
the fifth last snapshot. Therefore, our sample is complete for measuring
$f(\epsilon)$ except for the bin at $\epsilon=0.50$ where the
completeness is $98\%$ and the bin at $\epsilon=0.70$ with the completeness
of $97\%$.  

\begin{figure}
\begin{center}
\plotone{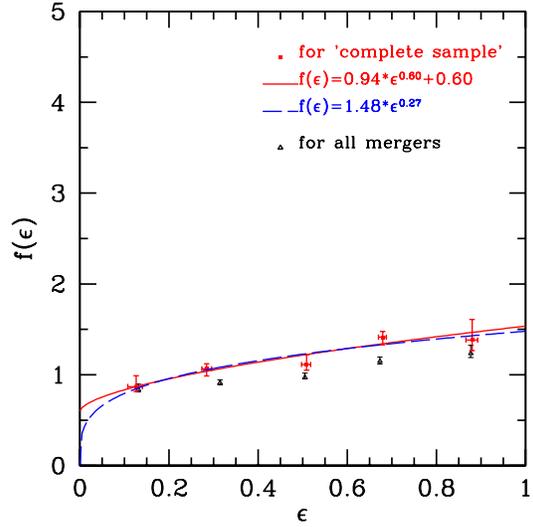}
\caption{\label{fig:func_circu_kang_c1_9} Fitting function of
$f(\epsilon)$.  $f(\epsilon)=0.94*{\epsilon}^{0.60}+0.60$ is
represented by the solid curve, while
$f(\epsilon)=1.48*{\epsilon}^{0.27}$ is denoted by the dashed
line. The squares are from the complete sample of mergers, and the
triangles are from all mergers.}
\end{center}
\end{figure}

In Figure~\ref{fig:func_circu_kang_c1_9} we present our estimate of
$f(\epsilon)$ from this complete sample. We first fit the data with
$f(\epsilon) = a \epsilon^\alpha$ and find the best fitting values
$a=1.48$ and $\alpha=0.27$. The fitting curve is displayed by the
dashed line.  
If we use all mergers identified instead of the complete sample, the
function $f(\epsilon)$ would be underestimated at larger $\epsilon$ as
shown by the triangles in the figure, which in turn would lead to an
even smaller $\alpha$ . However, the degree of the underestimation
becomes less serious, since the dependence on $\epsilon$ as shown by
the complete sample is much weaker than the original function
$f(\epsilon)=\epsilon^{0.78}$.   
To check if the circularity dependence
in Figure~\ref{fig:circu_dpd_kang_c1_8} is fully accounted for by this
fitting formula, we plot the median value of $T_{\rm simu}/T_{\rm
fit}$ as a function of the circularity parameter (dotted line in
Figure~\ref{fig:circu_dpd_kang_c1_8}). Compared with the dashed line
($f(\epsilon)= \epsilon^{0.78}$) the dependence on $\epsilon$ is
strongly reduced. However, we note that the new time ratio is still a
little higher for the smallest circularity bin. This can be
contributed to the artificial effect of the pure exponential fitting
form which falsely results in a vanishing merger timescale for
$\epsilon = 0$. If we consider two merging halos with equal masses and
assume they will merge within a free fall time scale $r_{\rm
vir}/V_{\rm c}$, equation $(4)$ gives $f(\epsilon)=0.60$. Therefore,
to avoid the artificial effect at $\epsilon=0$ due to the pure
exponential form and to reduce the somewhat too high time ratio in the
smallest circularity bin we fit our simulation data with $f(\epsilon)
= a \epsilon^\alpha+0.60$. The best fitting results are $a=0.94$ and
$\alpha =0.60$. The solid line in
Figure~\ref{fig:func_circu_kang_c1_9} shows the best fitting curve
which matches the data very well. The solid line in
Figure~\ref{fig:circu_dpd_kang_c1_8} demonstrates that the $T_ {\rm
simu}/T_{\rm fit}$ in the first circularity bin has moderately
decreased, now approaching to a value of 1.

The exponent $\alpha$ we find here is much smaller than the
widely used value $\alpha=0.78$. At this point it is worth recalling
that $\alpha=0.78$ was obtained by \cite{lacey93} analytically for the
case where a rigid satellite falls into an isothermal sphere. The fact
that our $f(\epsilon)$ always exceeds $\epsilon^{0.78}$ can be interpreted
as an indication for the mass loss of satellites in the simulation
\cite[cf.,][]{colpi99}. Satellites on radial orbits lose their mass
much faster than those on circular orbits, which implies that satellites
on the radial orbits show relatively prolonged merging time scales compared to
satellites on circular orbits. Therefore, one expects higher values
for $f(\epsilon)$ or equivalently lower values for $\alpha$ for small
$\epsilon$. In a future paper\citep[][in preparation]{fal07}, we will
explore this qualitative explanation using an analytical model similar
to those of \cite{zentner03} and \cite{zentner05} \citep[see
also,][]{bullock00, taylor01, taylor04}.

\subsection{Taking into account both the mass and circularity dependencies}
\label{sec:new}

\begin{figure}
\begin{center}
\plotone{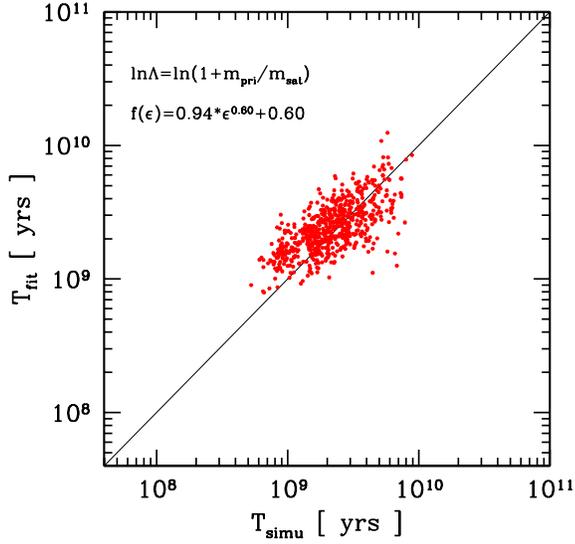}
\caption{\label{fig:merg_all_kang_c1_10} Comparison of the merger time
scale from simulation with our fitted merger time scale (equation
\ref{eq:new}). The data points lie slightly above the solid line,
because the sample missed a mall fraction of relatively long mergers.}
\end{center}
\end{figure}

Combining our results on the mass and circularity dependencies, we write 
the merger time scale as
\begin{equation}
\label{eq:new}
  T_{\rm fit}=\frac{0.94\epsilon^{0.60}+0.60}{2C}\frac {m_{\rm pri}}{m_{\rm sat}}
  \frac {1}{\ln [1+(\frac {m_{\rm pri}}{m_{\rm sat}})]}\frac{r_{\rm
vir}}{V_{\rm c}}\ .
\end{equation}
In Figure~\ref{fig:merg_all_kang_c1_10}  this equation is compared with
the merger time scale of all mergers measured in the
simulation. Remarkably, the scatter in the plot is much smaller than
that in Figures~\ref{fig:merg_all_ori_4}
and~\ref{fig:merg_all_kang_5}, indicating that $T_{\rm fit}$ describes
the merger timescale much better than equation $(1)$. 
To assess the scatter in more detail
Figure~\ref{fig:t_distri_kang_c1_11} displays the distribution of
$T_{\rm simu}/T_{\rm fit}$. The solid histogram shows the distribution
based on the early complete merger sample as described in \S
\ref{sec:revised_fe} and the dotted histogram gives the distribution
for all mergers identified in our simulation. 

The distribution for the sample of all mergers is shifted towards
the left relative to the complete sample. This is caused by the lack
of long time mergers among recently infalling satellites in the sample
of all mergers. These long time mergers would be included in the
sample if the simulation were evolved beyond the present time
$z=0$. This also leads to the trend of the data points to lie slightly
above the solid diagonal in Figure~\ref{fig:merg_all_kang_c1_10}.

The distribution of $x=T_{\rm simu}/T_{\rm fit}$ is well fitted
by the log-normal distribution
\begin{equation}
\label{eq:dis}
p(\ln x)d\ln x=\frac{1}{\sqrt{2\pi}\sigma}\exp\left[-\frac{(\ln x)^2}{2\sigma^2}\right]d \ln x
\end{equation}
with $\sigma=0.4$ (the smooth solid line in Figure
~\ref{fig:t_distri_kang_c1_11}). This distribution function combined
with the fitting function $(5)$ provides a description
for the merger time in a statistical sample.

From  Figure~\ref{fig:circu_dpd_kang_c1_8} we have learned
that the circularity dependence is accounted for by
equation $(5)$. Now, we examine the mass dependence when equation
$(5)$ is applied. Therefore, we plot the median value of $T_{\rm
simu}/T_{\rm fit}$ as a function of the mass ratio in
Figure~\ref{fig:mass_dpd_kang_c1_7} (solid line). Strikingly, 
we find no dependence on the mass ratio which indicates that 
the dependencies of the merger time on mass and circularity
are completely reproduced by equation $(5)$.

\subsection{Distribution of circularity}

Once the distribution of circularity is known for a population of
infalling satellites, one can determine how many of the satellites
will merge into central galaxies at a certain epoch in a statistical
way by using equations (\ref{eq:new}) and (\ref{eq:dis}).  Various authors
\citep[e.g.,][]{tormen97,zentner05,khochfar06} have studied this
distribution, with similar conclusions that orbits with intermediate
$\epsilon$ are common while those at both ends ($\epsilon\approx 0$ or 1) are rare.
Figure~\ref{fig:circu_para_all_12} shows our result for all resolved
satellite halos, with an average value of $\epsilon$ about $0.51$,
which is consistent with $0.53\pm 0.23$ in \cite{tormen97}. The
distribution can be well described by 
\begin{equation}
\label{eq:cdis}
p(\epsilon)d\epsilon=2.77{\epsilon}^{1.19}(1.55-\epsilon)^{2.99}d\epsilon
\end{equation}
which is shown as the solid line in Figure
\ref{fig:circu_para_all_12}.  The circularity distribution is
independent of the mass ratio of the primary halo and the satellite,
as shown by Figure~\ref{fig:circu_mratio_all_13}.

It is worth noting that in semi-analytical models, the circularity
parameter was often randomly drawn from a uniform distribution between
$0$ and $1$ \citep[e.g.,][]{kauffmann99,somerville99}. According
to our findings such an approach biases the estimate for the input
dynamical friction timescales. 

\section{Conclusions and discussion}

In this paper, we have analysed galaxy mergers in a SPH/N-body 
simulation and compared the merger time scale with the theoretical
prediction based on the Chandrasekhar formula. We have obtained the
following main conclusions. 
\begin{itemize}
\item In contrast with \cite{navarro95}, we find that the widely used
equation (\ref{eq:dynf}) with the satellite's total mass at its first
crossing of the host virial radius taken for $m_{\rm sat}$,
systematically underestimates the merger timescale for minor mergers
and overestimates it for major mergers; 
\item We show that the two alternative forms $\ln (1+m_{\rm
pri}/m_{\rm sat})$ and $1/2\ln [1+(m_{\rm pri}/m_{\rm sat})^2]$ for
the Coulomb logarithm, which also are widely used in literature,
account for the mass dependence of merger timescale
successfully. However, both of them underestimate the merger time
scale by a factor 2 if the satellite's total mass at its first
crossing of the host virial radius is used for $m_{\rm sat}$. Of these
two forms, the former does slightly better in accounting for the mass
dependence; 
\item With $\ln (1+m_{\rm pri}/m_{\rm sat})$ taken for the Coulomb
logarithm, we find the dependence on circularity parameter $\epsilon$
is much weaker than $\epsilon^{0.78}$, and can be accurately
represented by $ 0.94{\epsilon}^{0.60}+0.60$;
\item Combining our findings on the mass and circularity dependencies,
we present an accurate fitting formula (eq.\ref{eq:new}) for
the merger timescale. Together with the distribution functions
(eqs.\ref{eq:dis} and \ref{eq:cdis}), one can use this equation to
predict for mergers of galaxies in LCDM models.
\end{itemize}

Our results do not necessarily mean that Chandrasekar's theory
is not applicable for mergers of galaxies. Instead our results do indicate
that many previous applications of this theory led to incorrect results 
because some simplified assumptions were adopted. We believe
that the mass loss of satellites and the steep density gradient of
host halos are two of the key reasons that make the problem complicated. 
In a future paper, we will investigate if our simulation results can be
reproduced with the Chandrasekhar theory by properly taking into
account of these two factors.

In the following we will discuss how potential shortcomings in the
treatment of the baryonic physics at the core of the primary halo may
affect our results. It is well known that  current hydrodynamic
simulations suffer from the so called 'overcooling' problem, i.e. the
gas at the core of massive dark matter halos cools too rapidly
resulting in too massive central galaxies compared to observations 
\citep[e.g.,][]{borgani04,saro06,naab07}. In turn, adiabatic contraction
\citep[e.g.,][]{gnedin04} may also change the dark matter properties
at the central parts of the halo in an unphysical manner.    

\begin{figure}

\begin{center}
\plotone{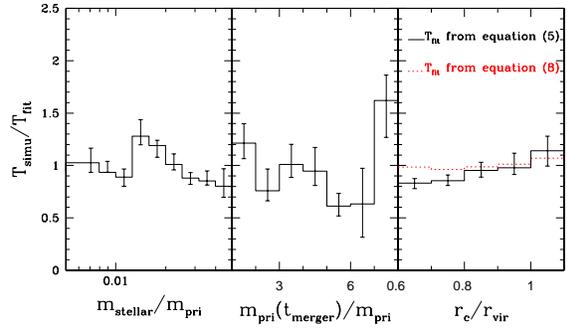}
\caption{\label{fig:tratio} The ratio $T_{\rm simu}/T_{\rm fit}$ as a
function of the mass ratio $m_{\rm stellar}/m_{\rm pri}$ (left), the
growth rate of the primary halo (middle), and $r_c/r_{\rm vir}$ (right)
for the complete sample. }
\end{center}
\end{figure}

However, we think that this process does not substantially alter our
results for two reasons: (1) With exception of very radial orbits,
which are rare, the satellite galaxies spend most of their time during
the merging process at the outer parts of the primary halo where
dynamical friction is moderate. If, however, the satellite is once
migrated towards the central parts of the primary halo dynamical
friction becomes very efficient and the remaining lifetime of the
satellite galaxy is short. Consequently, the merger time scale is set
by the conditions at the outer parts of the primary halo
\citep[c.f.,][]{navarro95}. (2) The findings of \cite{springel01} and
\cite{kang05} are in qualitative agreement with our results. Since
both of these studies are based on pure N-body simulations they
obviously do not suffer from the overcooling problem. This is an
further indication that our results are accurate and are not affected
by the potential shortcomings in the treatment of the baryonic physics
in the simulations.  These arguments are supported by the left panel
of Figure \ref{fig:tratio}, which shows that there is no dependence of
$T_{\rm simu}/T_{\rm fit}$ on $m_{\rm stellar}/m_{\rm pri}$, the ratio
of the central galaxy's stellar mass to the dark matter mass of the
surrounding primary halo.

We have also checked if our result is affected by the growth of the
primary halo during the merger course. The middle panel of Figure
\ref{fig:tratio} shows the ratio $T_{\rm simu}/T_{\rm fit}$ as a
function of the mass growth rate of the primary halo, which is defined
as the ratio of its dark matter mass at the time of merger to its mass
at the time of the first crossing. The result indicates that the
merger time scale is not affect by the growth of the primary halo. A
possible explanation is that the internal density structure does
not change significantly during this course. Of course, violent major
mergers may change the internal structures and bring about large
fluctuations in the merger time.

To keep the fitting formula simple to use, we prefer not to include
the dependence on the energy of the satellite's orbit, that is, on
$r_c$. We have examined this dependence in the right panel of
Figure \ref{fig:tratio}, which shows there is a weak dependence on
$r_c/r_{\rm vir}$. We can include this dependence in our fitting
formula by replacing $r_{\rm vir}$ with $\sqrt{r_{\rm vir}
r_c}$. Thus, the fitting formula with the $r_c$-dependence included
reads as 
\begin{equation}
\label{eq:newrc}
  T_{\rm fit}=\frac{0.90\epsilon^{0.47}+0.60}{2C}\frac {m_{\rm
  pri}}{m_{\rm sat}} \frac {1}{\ln [1+(\frac {m_{\rm pri}}{m_{\rm
  sat}})]}\frac{\sqrt{r_{\rm vir}r_c}}{V_{\rm c}}\ ,
\end{equation}
and Figure\ref{fig:tratio} shows that the $r_c$ dependence is fully
accommodated by this simple heuristic correction. We have checked the
dependence on the mass ratio $m_{\rm pri}/m_{\rm sat}$ as well as the
scatter in $T_{\rm simu}/T_{\rm fit}$, and found that they are nearly
the same as when equation (\ref{eq:new}) is used. The better
performance of equation (\ref{eq:newrc}) is achieved at the expense of
computing the energy of the individual satellite orbits. 
Since the accuracy of the fitting formula is improved only
slightly by including the $r_c$ dependence compared with the scatter
in $T_{\rm simu}/T_{\rm fit}$, the simpler formula (\ref{eq:new}) is
preferred for most applications.

As concluding remark, we once again focus our attention on equation  
(\ref{eq:new}) which can be considered as the distillate of 
our analysis. This fitting formula allows to predict the merger timescale
$T_{\rm fit}$ for the two central galaxies within a satellite and a
primary halo. The merger timescale for the satellite  galaxy is
defined as the time which elapses between its  first crossing of the
primary's virial radius and its final  coalescence with the central
galaxy.  The computation of accurate merger timescales is a  crucial
ingredient for semi-analytical modeling of galaxy  formation. 

\begin{figure}
\begin{center}
\plotone{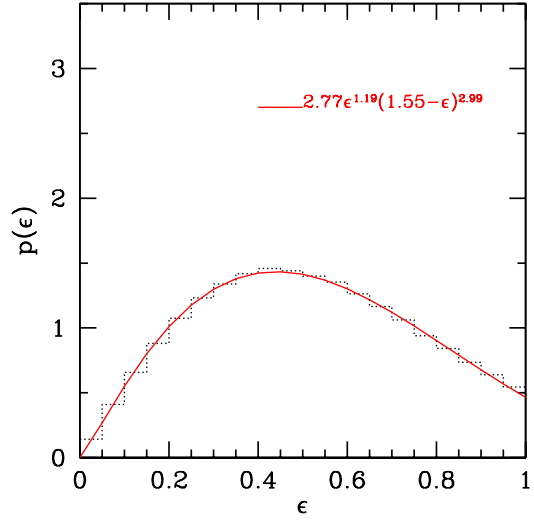}
\caption{\label{fig:circu_para_all_12} Distribution of circularity
parameter $\epsilon$ for all resolved satellite halos having more
than half their masses entering the primary halos.}
\end{center}
\end{figure}

\begin{figure}
\begin{center}
\plotone{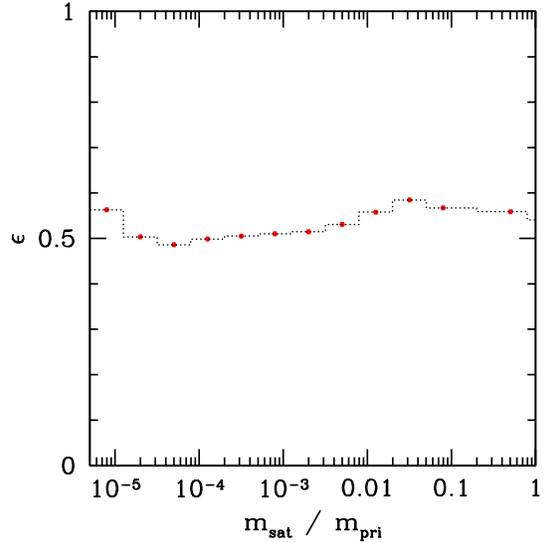}
\caption{\label{fig:circu_mratio_all_13} The mean circularity as a
function of the mass ratio.}
\end{center}
\end{figure}

Equation (\ref{eq:new}) requires two input values: $m_{\rm pri}/m_{\rm
sat}$, the mass ratio of primary and satellite halo (before they start
merging) and $\epsilon = J/J(E)$, the satellites initial circularity
which is defined as the ratio of the satellites actual angular
momentum $J$ and the angular of a circular orbit with the same energy
$J(E)$. $r_{\rm vir}$ is the virial radius of the primary halo just
before the satellite merges with it. The factors $C$ and $r_{\rm
vir}/V_c$ are constants and do not depend on the specific
constellation. If $\epsilon$ is not known it can be randomly drawn
from the distribution provided by equation (\ref{eq:cdis}) which we
have derived directly from the simulation data, see
Figure~\ref{fig:circu_para_all_12}. This random process can be applied
for arbitrary mass ratios ($m_{\rm pri}/m_{\rm sat}$) since the
distributions of $\epsilon$ are nearly independent of mass as shown in
Figure~\ref{fig:circu_mratio_all_13}. 

\begin{figure}
\begin{center}
\plotone{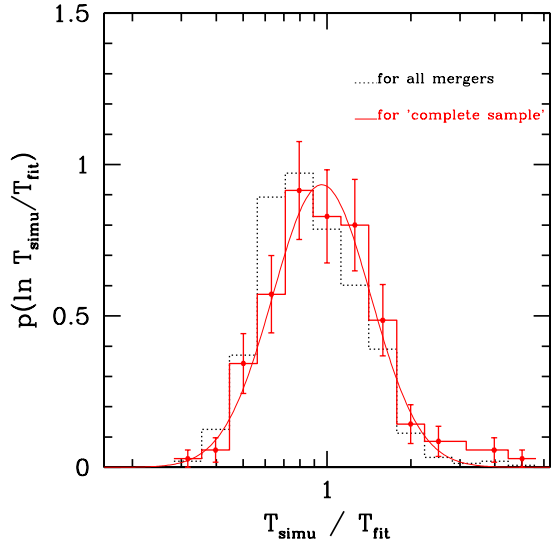}
\caption{\label{fig:t_distri_kang_c1_11} Distribution of $T_{\rm
simu}/ T_{\rm fit}$ for the whole merger sample(the dotted line) and
for the complete sample(the solid line). The left shift of the whole
merger sample is mainly due to its lack of those relatively long
mergers.}
\end{center}
\end{figure}

Finally, it remains to mention
that due to stochastic processes during a merger event, like
close encounters with other substructures or the occurrence of
multiple mergers at the same time, there arises substantial scatter
among the merger time scales with equivalent initial conditions. This
can be taken into account if the values for $T_{\rm fit}$ obtained from
equation (\ref{eq:new}) are spread according to the log-normal
distribution given in equation (\ref{eq:dis}) which is also
displayed in Figure~\ref{fig:t_distri_kang_c1_11}. 
With the fitting formula (\ref{eq:new}) we provide a robust
estimate of the merger timescale pivotal for all kinds of
analytical modeling of galaxy evolution within dark matter
halos. 

After we submitted our paper both to the journal and to the astro-ph
electronic library, an independent work by \citet{b-k07} on the same
subject had appeared on the electronic library.  Their paper is
qualitatively consistent with ours in that the time scale given by
equation (\ref{eq:dynf}) is underestimated.  But quantitatively, their
results are rather different from ours. Both the dependencies on the
mass ratio and the circularity parameters are much stronger in their
paper. In particular, the strong dependence on the circularity they
found, which is even stronger than $\epsilon^{0.78}$ at
$\epsilon=0.5-1$, is not consistent with our data. The dependence on
the mass ratio is also stronger than ours. This discrepancy may mainly
come from the difference between the simulations: they present a series of
pure N-body simulations of two halo mergers, whereas our results are based on a
cosmological hydro/N-body simulation with star formation. 
While our fitting formula is accurate for mergers in the cosmological
frame, future work is still needed to find out the specific causes of this
discrepancy.

  \acknowledgments We would like to thank Volker
Springel for providing the Gadget Code with star formation and Liang
Gao for his help with using the code.  This work is supported by NSFC
(10533030, 0742961001, 0742951001), by Shanghai Key Projects in Basic
research (No. 04JC14079 and 05XD14019), and by the Knowledge
Innovation Program of the Chinese Academy of Sciences, Grant
No. KJCX2-YW-T05. AF is supported by the CAS Research Fellowship for
International Young Researchers. The simulation was performed at the
Shanghai Supercomputer Center.


\end{document}